\pdfoutput=1
\RequirePackage{ifpdf}
\ifpdf % We~are running pdfTeX in pdf mode
\documentclass[pdftex]{sigma}%,draft
\else
\documentclass{sigma}
\fi

\numberwithin{equation}{section}
\newtheorem{Theorem}{Theorem}[section]
\newtheorem{Corollary}[Theorem]{Corollary}

 { \theoremstyle{definition}

 }

\def\Re{\operatorname{Re}}
\def\Im{\operatorname{Im}}

\begin{document}
\allowdisplaybreaks

\newcommand{\arXivNumber}{2106.03835}

\renewcommand{\thefootnote}{}

\renewcommand{\PaperNumber}{091}

\FirstPageHeading

\ShortArticleName{Negative Times of the Davey--Stewartson Integrable Hierarchy}

\ArticleName{Negative Times of the Davey--Stewartson\\ Integrable Hierarchy\footnote{This paper is a~contribution to the Special Issue on Mathematics of Integrable Systems: Classical and Quantum in honor of Leon Takhtajan.

The full collection is available at \href{https://www.emis.de/journals/SIGMA/Takhtajan.html}{https://www.emis.de/journals/SIGMA/Takhtajan.html}}}

\Author{Andrei K.~POGREBKOV~$^{\rm ab}$}

\AuthorNameForHeading{A.K.~Pogrebkov}

\Address{$^{\rm a)}$~Steklov Mathematical Institute of Russian Academy of Sciences, Moscow, Russia}
\EmailD{\href{mailto:pogreb@mi-ras.ru}{pogreb@mi-ras.ru}}
\Address{$^{\rm b)}$~Skolkovo Institute of Science and Technology, Moscow, Russia}

\ArticleDates{Received June 08, 2021, in final form October 01, 2021; Published online October 12, 2021}

\Abstract{We use example of the Davey--Stewartson hierarchy to show that in addition to the standard equations given by Lax operator and evolutions of times with positive numbers, one can consider time evolutions with negative numbers and the same Lax operator. We~derive corresponding Lax pairs and integrable equations.}


\Classification{37K10; 70H06}

\begin{flushright}
 \it To Leon A.\ Takhtajan with best wishes
\end{flushright}

\renewcommand{\thefootnote}{\arabic{footnote}}
\setcounter{footnote}{0}

\section{Introduction}
In \cite{akp2008a}, we proposed a method of derivation of $(2+1)$-dimensional nonlinear integrable equations based on commutator identities on associative algebras. Taking into account the algebraic similarity of operator commutators and derivatives, we have transformed commutator identities into linear partial differential equations. A characteristic property of these linear equations is the possibility to lift them up to nonlinear, integrable ones. In \cite{akp2008b, akp2010}, this approach was extended to differential-difference and difference equations, where the analogy of similarity transformations and shifts of independent variables was used. In~\cite{akp2016}, we developed this result for non-Abelian identities of commutators.

To formulate the main aspects of this approach, we start here with the simplest examples. Let $A$ and $B$ be arbitrary elements of an arbitrary associative algebra~$\mathcal{A}$. Then they obey the commutator identity
\begin{gather}
4\big[A_{}^{3},[A,B]\big]-3\big[A_{}^{2},\big[A_{}^{2},B\big]\big]-[A,[A,[A,[A,B]]]]=0.\label{commut1}
\end{gather}
Being a trivial consequence of associativity, this identity easily proves that the function
\begin{gather}\label{Bt1}
B(t_1, t_2,t_3)={\rm e}^{t_1A+t_2A^2+t_3A^3}_{}B{\rm e}^{-t_1A-t_2A^2-t_3A^3},
\end{gather}
i.e., such that $B_{t_n}=[A^{n}, B]$, $n=1,2,3$, obeys the linearized Kadomtsev--Petviashvili (KP)\footnote{More precisely, KPII.} equation with respect to the variables $t_j$:
\begin{gather*}
 4\dfrac{\partial_{}^{2}B(t)}{\partial t_{1}^{}\partial t_{3}^{}}-
 3\dfrac{\partial_{}^{2}B(t)}{\partial t_{2}^{2}}-
 \dfrac{\partial_{}^{4}B(t)}{\partial t_{1}^{4}}=0.
\end{gather*}
It was stated in~\cite{akp2008a,akp2020} that there are similar relations for higher commutators. In the case of KP, they lead to higher linear equations\vspace{-1ex}
\begin{gather*}
2^{n}\partial_{t_n}^{}\partial_{t_1}^{n}B
=\bigl(\partial_{t_2}^{}+\partial_{t_1}^{2}\bigr)^{n}B-\bigl(\partial_{t_2}^{}-\partial_{t_1}^{2}
\bigr)^{n}B, \qquad n=3,4,\dots.
\end{gather*}

Similar results were obtained in \cite{akp2008b,akp2010,akp2016} for difference and differential-difference equations. In that case, we replace (\ref{commut1}) with, say, the commutator identity\vspace{-1ex}
\begin{gather*}
\big[A, \big[A^{-1}, B\big]\big] = 2B-ABA^{-1} - A^{-1}BA,
\end{gather*}
where the element $A$ is assumed to be invertible. Thus, in addition to commutators of the kind~(\ref{commut1}), we get similarity transformations here (commutators in the group sense). Therefore, we~introduce the element $B$ depending on the number $n_1$ and continuous variables $t_1$ and $t_{-1}$ by means of\vspace{-1ex}
\begin{gather}\label{Bt2}
 B(t_1,t_{-1},n_1)={\rm e}^{t_1A+t_{-1}A^{-1}}_{}A^{n_1}BA^{-n_1}{\rm e}^{-t_1A-t_{-1}A^{-1}}_{},
\end{gather}
and denote the shift with respect to variable $n_1$ as $B^{(1)}=ABA^{-1}$. Accordingly, this element $B$ obeys the linear differential-difference equation
\begin{gather*}
B_{t_1,t_{-1}}=2B-B^{(1)}-B^{(-1)},
\end{gather*}
which gives a linearized version of the two-dimensional Toda system \cite{T1, T2}.

In \cite{akp2008a,akp2008b,akp2010}, we proved that any linear equation, resulting from the commutator identity, can be lifted up to a nonlinear integrable equation using a special dressing procedure. In this paper, our goal is to extend the class of commutator identities. For this purpose one can use arbitrary functions $f(A)$ with commutativity being the only condition they should obey: $[f(A),g(A)]=0$. A natural
generalization of the choice of functions of the element $A$ was suggested in \cite{akp2008a,akp2020}. We~assume that in algebra $\mathcal{A}$ there exists an element $\sigma$ such that\vspace{-1ex}
\begin{gather}
\sigma^{2}=1,\qquad[A,\sigma]=0, \qquad\{B,\sigma\}=0,\qquad [A\sigma,B]=\sigma\{A,B\},\label{init}
\end{gather}
where $\{\cdot,\cdot\}$ denotes anticommutator. In particular, we can consider elements of $\mathcal{A}$ as $2\times2$ matrices, where $A$ is proportional to the unity matrix $I$, $B$ is off diagonal,\vspace{-1ex}
\begin{gather}
B=\begin{pmatrix}	0 & B_1	\\ B_2 & 0
\end{pmatrix}\!,\label{init2}
\end{gather}
and $\sigma=\sigma_3$ is a Pauli matrix. Thus the commutator of $A\sigma$ and $B$ reduces to the anticommutator of $A$ and $B$, so that for any $n$ we have the commutator identity\vspace{-1ex}
\begin{gather*}
\big[A^{2n},B\big]=\sigma\big[A^{n},\big[\sigma A^{n},B\big]\big].
\end{gather*}

We consider the commutators $[A,B]$ and $[A\sigma,B]$ as generating and decompose the commutators $[A^n,B]$ or $[A^n\sigma,B]$ for $n\geq2$ in their terms. Thanks to (\ref{init}), (\ref{init2}) it is easy to prove the following commutator identities:
\begin{gather}
2^{n}\sigma^{n+1}\big[A^{n},B\big]= \underbrace{[(\sigma+I)A,[(\sigma+I)A,\dots,[(\sigma+I)A}_n,B],\dots]\nonumber
\\[-1ex]
\hphantom{2^{n}\sigma^{n+1}\big[A^{n},B\big]=}{} -\underbrace{[(\sigma-I)A,[(\sigma-I)A,\dots,[(\sigma-I)A}_n,B],\dots],\label{higher4}
\\
2^{n}\sigma^{n+1}\big[\sigma A^{n},B\big]= \underbrace{[(\sigma+I)A,[(\sigma+I)A,\dots,[(\sigma+I)A}_n,B],\dots]\nonumber
\\[-1ex]
\hphantom{2^{n}\sigma^{n+1}\big[\sigma A^{n},B\big]=}{} +\underbrace{[(\sigma-I)A,[(\sigma-I)A,\dots,[(\sigma-I)A}_n,B],B],\dots],\label{higher5}
\end{gather}
where $n\geq1$. These two sets of commutator identities give two sets of differential hierarchies if, in addition to (\ref{Bt1}), we introduce two sets of variables, $t=\{t_1,t_2,\dots\}$ and $x=\{x_1,x_2,\dots\}$, given by the equations
\begin{subequations}\label{comm}
\begin{gather}
B_{t_n}^{}=\big[A^{n},B\big],\label{comm1}
\\
B_{x_n}^{}=\big[\sigma A^{n},B\big].\label{comm1'}
\end{gather}
\end{subequations}
Taking $n=1$ here, we get
\begin{gather}
(\partial_{x_1}\pm\partial_{t_1})^{n}B=\underbrace{[(\sigma\pm I)A,\dots,[(\sigma\pm I)A}_n,B],\dots],\label{comm0}
\end{gather}
so thanks to (\ref{higher4}) and (\ref{higher5}), we get linear differential equations for $B(t,x,z)$,
\begin{gather}
2^{n}\sigma^{n+1}\partial_{t_n}B=(\partial_{x_1}+\partial_{t_1})^{n}B-(\partial_{x_1}-\partial_{t_1})^{n}B,
\label{i12}	
\\
2^{n}\sigma^{n+1}\partial_{x_n}B=(\partial_{x_1}+\partial_{t_1})^{n}B+(\partial_{x_1}-\partial_{t_1})^{n}B.
\label{i13}
\end{gather}
For $n=2$, these equalities are read as $\sigma B_{t_2}=B_{t_1x_1}$ and $\sigma B_{x_2}=B_{t_1t_1}+B_{x_1x_1}$, respectively. In \cite{akp2020}, these linear equations were lifted to the Davey--Stewartson equation (see \cite{DS}) and higher equations of its hierarchy.

Here we consider ``negative'' version of this hierarchy, i.e., we assume negative values of $n$ in~(\ref{comm}). In Section~\ref{sec2}, we derive the corresponding commutator identities and the corresponding linear differential equations. In Section~\ref{sec3}, we introduce the realization of elements of the associative algebra $\mathcal{A}$ using pseudo-differential operators. On this basis, in Section~\ref{sec4}, we consider the dressing procedure that enables introduction of the dressing operator and its time evolutions. The Lax pair and nonlinear equations are derived in Section~\ref{sec5}. Section~\ref{sec6} is devoted to $(1+1)$-dimensional reductions of the systems under consideration. Some concluding remarks are given in Section~\ref{sec7}.

\section{Commutator identities and linear equations}\label{sec2}
Our goal here is to construct a commutator identity based on the commutators $[A,B]$, $[\sigma A,B]$, and $\big[A^{-1},B\big]$ or $[A,B]$, $[\sigma A,B]$, and $\big[\sigma A^{-1},B\big]$, where we assume existence of the inverse element~$A^{-1}$. By analogy with the above, we consider commutators $[A,B]$ and $[\sigma A,B]$ as generating for the commutators $\big[A^{-1},B\big]$ and $\big[\sigma A^{-1},B\big]$. It is easy to check that we have here the following commutator identities:
\begin{gather}
\big[\sigma A,\big[\sigma A,\big[A^{-1},B\big]\big]\big]-\big[A,\big[A,\big[A^{-1},B\big]\big]\big]+4[A,B]=0,
\label{commut3}
\\
\big[\sigma A,\big[\sigma A,\big[\sigma A^{-1},B\big]\big]\big]-\big[A,\big[A,\big[\sigma A^{-1},B\big]\big]\big]-4[\sigma A,B]=0.
\label{commut4}
\end{gather}
Taking into account that all these commutators mutually commute, we consider $B$ as a function of $t_1$, $x_1$ and $t_{-1}$, or $x_{-1}$, such that
\begin{subequations}\label{comm2}
\begin{gather}
 B_{t_1}=[A,B],\qquad B_{x_1}=[\sigma A,B],\label{comm21}
\\
B_{t_{-1}}=\big[A^{-1},B\big],\qquad B_{x_{-1}}=\big[\sigma A^{-1},B\big].\label{comm22}
\end{gather}
\end{subequations}
Thanks to (\ref{comm2}), we get from (\ref{commut3}) and (\ref{commut4}) linear equations of motion (cf.\ (\ref{i12}) and (\ref{i13}) for~$n=2$)
\begin{gather}
B_{x_1x_1t_{-1}}-B_{t_1t_1t_{-1}}+4B_{t_1}=0,\label{ci1}
\\
B_{x_1x_1x_{-1}}-B_{t_1t_1x_{-1}}-4B_{x_1}=0.\label{ci2}
\end{gather}
Thus, we again have two versions of the equations: one involving $\partial_{t_{-1}}$, and the other invol\-ving~$\partial_{x_{-1}}$. Taking into account the symmetry of these two equations with respect to the substitution $x_{-1}\leftrightarrow-t_{-1}$, we study here mainly (\ref{ci1}).

By extending (\ref{comm}) to negative values of $n$, we arrive at a hierarchy of commutator identities and linear equations. We can use (\ref{i12}) and~(\ref{i13}), substitute $n\to-n$ into these equations and multiply them both by $\big(\partial_{x_1}^{2}-\partial_{t_1}^{2}\big)$. Thus, we get
\begin{gather}
\bigl(\partial_{t_{-n}}\big(\partial_{x_1}^{2}-\partial_{t_1}^{2}\big)^{n}+2^{n}(\sigma\partial_{x_1}+\partial_{t_1})^{n}-
2^{n}(\sigma\partial_{x_1}-\partial_{t_1})^{n}\bigr)B=0,
\label{ci3}
\\
\bigl(\sigma\partial_{x_{-n}}\big(\partial_{x_1}^{2}-\partial_{t_1}^{2}\big)^{n}B-2^{n}(\sigma\partial_{x_1}+
\partial_{t_1})^{n}-2^{n}(\sigma\partial_{x_1}-\partial_{t_1})^{n}\bigr)B=0,
\label{ci4}
\end{gather}
where $n=1,2,\dots$ and where by analogy with (\ref{comm22})
\begin{gather}
B_{t_{-n}}=\big[A^{-n},B\big],\qquad B_{x_{-n}}=\big[\sigma A^{-n},B\big].\label{comm3}
\end{gather}
We omit here form of (\ref{ci3}) and (\ref{ci4}) in terns of commutator identities. It can be easely restored with the help of (\ref{comm0}). In the case of $n=1$, equations (\ref{ci3}) and (\ref{ci4}) are reduced to (\ref{commut3}) and~(\ref{commut4}). Now we have to show that all these linear equations admit lift up to nonlinear integrable ones.

\section{Realization of elements of the associative algebra}\label{sec3}

To this end, we consider a special realization of the elements of the associative algebra $\mathcal{A}$, see \cite{akp2008a, akp2008b, akp2010, akp2016}. By analogy with the standard definition of the pseudo-differential operators, we define an element $F$ of $\mathcal{A}$ by its symbol $\widetilde{F}(t,x,z)$. Here $t$ and $x$ denote (finite) subsets of real variables $t=\{\dots,t_{-2},t_{-1},t_1,t_2\dots\}$, $x=\{\dots,x_{-2},x_{-1},t_1,t_2,\dots\}$, and $z\in\mathbb{C}$ denotes a~complex parameter. The subsets $t$ and $x$ definitely include the variables $t_1$ and $x_1$ and at least one of the other variables of these lists. In the following we call such subsets \textit{minimal}. The symbol of the composition of two elements of the algebra is given by means of the symbols of cofactors in the form
\begin{gather}
\widetilde{FG}(t,x,z)=\dfrac{1}{2\pi}\int {\rm d}p\int {\rm d}y\,\widetilde{F}(t,x,z+{\rm i}p){\rm e}^{{\rm i}p(t_1-y)}\widetilde{G}(y,t',x,z),\label{comp}
\end{gather}
where $t'$ denotes a subset $t$ without variable $t_1$. We see that the variable $t_1$ plays a special role here: the composition with respect to other variables is pointwise. In what follows we consider elements of the algebra $\mathcal{A}$ such that their symbols belong to the space of tempered distributions of their arguments. The symbol of the unity operator is $1$, and we choose the symbol of operator~$A$~as
\begin{gather}
\widetilde{A}(t,x,z)=z.\label{Az}
\end{gather}
Thanks to (\ref{comp}) we have that for any $F$
\begin{gather*}
\widetilde{A^nF}(t,x,z)=(z+\partial_{t_1})^{n}\widetilde{F}(t,x,z),\qquad
\widetilde{FA^n}(t,x,z)=z^{n}\widetilde{F}(t,x,z),
\end{gather*}
where $A^n$ is understood as $n$-th power of composition (\ref{comp}), where now $n\in\mathbb{Z}$. Then, for $n=1$, we get $[A,F]=\partial_{t_1}F$ according to (\ref{comm1}). Further relations of these equalities give in terms of symbols:
\begin{gather*}
\widetilde{B}_{t_n}^{}(t,x,z)=\bigl((z+\partial_{t_1})^{n}-z^{n}\bigr)\widetilde{B}(t,x,z),
\\
\widetilde{B}_{x_n}^{}(t,x,z)=\sigma\bigl((z+\partial_{t_1})^{n}+z^{n}\bigr)\widetilde{B}(t,x,z).
\end{gather*}

Because of our assumption, the symbol $\widetilde{B}(t,x,z)$ admits a Fourier transform with respect to the variable $t_1$, so the above relations show
\begin{gather}
\widetilde{B}(t,x,z)=\int {\rm d}p\,\exp\biggl(\sum_{n}\bigl((z+{\rm i}p)^{n}-z^{n}\bigr)t_n
+\sigma\sum_{n}\bigl((z+{\rm i}p)^{n}+z^{n}\bigr)x_n\biggr)f(p,z),
\label{Bf}
\end{gather}
where $n\in\mathbb{Z}$ and $f(p,z)$ is an arbitrary $2\times2$ off diagonal matrix function independent of all~$t_n$ and $x_n$. Note that here we do not specify set of ``times'' $t_i$ and $x_i$ involved in the evolution equation. We know that this set includes at least three times: $t_1$, $x_1$ and one of times $t_n$, or~$x_n$ with $n\neq0$ and $1$. It can include more times, but $t_1$ and $x_1$ and every third time gives an evolution equation generated by the commutator identity. Thus, in (\ref{Bf}), summation in the exponent goes over finite number of terms, corresponding to times that are ``switched on'' while other times are equal to zero.

It is natural to impose on $\widetilde{B}(t,x,z)$ the conditions of convergence of the integral and the boundedness of the limits of $\widetilde{B}(t,x,z)$ as $t$, or $x$ tends to infinity. Two obvious conditions are sufficient for this. The first one is given by the choice
$f(p,z)=\delta(p+2z_{\Im})g(z)$, where $\delta$ denotes delta-function, so that (\ref{Bf}) takes the form
\begin{gather}
\widetilde{B}(t,x,z)=\exp\biggl(\sum_{n}\bigl(\overline{z}^{n}-z^{n}\bigr)t_n
+\sigma\sum_{n}\bigl(\overline{z}^{n}+z^{n}\bigr)x_n\biggr)g(z),\label{Bf1}
 \end{gather}
where $g(z)$ is an arbitrary bounded function of its argument. But in order to get $\widetilde{B}(t,x,z)$ bounded with respect to variables $x_n$, it is necessary to perform substitution
\begin{gather}
x_n\to {\rm i}x_n,\label{bf1}
\end{gather}
where the new $x_n$ are real.

The second case is given by reduction $f(p,z)=\delta(z_{\Re})h(p,z_{\Im})$, where $z=z_{\Re}+{\rm i}z_{\Im}$ and~$h(p,z_{\Im})$ is an arbitrary function of its arguments. Then (\ref{Bf}) takes the form
\begin{gather}
 \widetilde{B}(t,x,z)=\int {\rm d}p\,\exp\biggl(\sum_{n} {\rm i}^{n}\bigl((z_{\Im}+p)^{n}-z_{\Im}^{n}\bigr)t_n
 +\sigma {\rm i}^n\sum_{n}\bigl((z_{\Im}+p)^{n}+z_{\Im}^{n}\bigr)x_n\biggr)\nonumber
 \\ \hphantom{\widetilde{B}(t,x,z)=\int {\rm d}p}
 {}\times h(p,z_{\Im})\delta(z_{\Re}),\label{Bf2}
\end{gather}

Here we see that $\widetilde{B}(t,x,z)$ is bounded with respect to variables $t_n$ and $x_n$ with odd numbers, and in order to make it bounded for
variables with even numbers, we need to make a substitution
\begin{gather}
t_{2n}\to {\rm i}t_{2n},\qquad x_{2n}\to {\rm i}x_{2n}.\label{bf2}
\end{gather}
Thus we have two types of systems defined by the choices (\ref{Bf1}) and (\ref{Bf2}).

\section{Dressing procedure}\label{sec4}
Specific property of the above set of operators is the possibility of defining operation of $\bar\partial$-differentiation by the complex variable $z$, $F\to\bar\partial{F}$. In terms of symbols, this is defined, see~\cite{akp2008a},~as
\begin{gather}
\big(\widetilde{\bar{\partial}F}\big)(t,x,z)=\dfrac{\partial {\widetilde{F}(t,x,z)}}{\partial\overline{z}},\label{dbar}
\end{gather}
{\samepage
where derivative is understood in the sense of distributions. Thanks to (\ref{Az}), we get the equality
\begin{gather}
\bar{\partial}A=0,\label{A0}
\end{gather}
which plays essential role in what follows.

}

Now we can define a dressing operator $K$ with symbol $\widetilde{K}(t,x,z)$ by means of $\overline\partial$-problem
\begin{gather}
\overline\partial K=KB,\label{dbar1}
\end{gather}
where the product in r.h.s.\ is understood in the sense of the composition law (\ref{comp}). Thanks to~(\ref{comp}) and (\ref{dbar}), the equality (\ref{dbar1}) takes the explicit form
\begin{gather}
\dfrac{\partial\widetilde{K}(t,x,z)}{\partial\overline z}=
\widetilde{K}(t,x,\overline{z})\exp\biggl(\sum_{n}\bigl(\overline{z}^{n}-z^{n}\bigr)t_n
+\sigma\sum_{n}\bigl(\overline{z}^{n}+z^{n}\bigr)x_n\biggr)g(z),
\label{Bf1expl}
\end{gather}
for time evolutions given by (\ref{Bf1}) and the form
\begin{gather}
\dfrac{\partial\widetilde{K}(t,x,z)}{\partial\overline z}=\delta(z_{\Re})\int {\rm d}p\,\widetilde{K}(t,x,{\rm i}p)\times\nonumber
\\ \hphantom{\dfrac{\partial\widetilde{K}(t,x,z)}{\partial\overline z}=}
{}\times\exp\biggl(\sum_{n}{\rm i}^{n}\bigl((p^{n}-z_{\Im}^{n})t_n
+\sigma (p^{n}+z_{\Im}^{n})x_n\bigr)\biggr)h(p-z_{\Im},z_{\Im}),\label{Bf2expl}
\end{gather}
for time evolutions given by (\ref{Bf2}). Thus, in the case of (\ref{Bf1expl}), the equation (\ref{dbar1}) gives the $\overline\partial$-problem, while in the case of (\ref{Bf2expl}) we get Riemann--Hilbert problem. In both these cases, we normalize solution $K$ of the equation (\ref{dbar1}) by the asymptotic condition
\begin{gather}
\widetilde{K}(t,x,z)\to 1,\qquad
z\to\infty.\label{dbar2}
\end{gather}

In what follows, we assume unique solvability of the problem (\ref{dbar1}), (\ref{dbar2}). The time evolution of the dressing operator follows from these equations. Say, due to (\ref{comm}) and (\ref{comm2}) we get
\begin{gather}
\overline\partial K_{t_n}=K_{t_n}B+K\big[A^{n},B\big],\qquad
\overline\partial K_{x_n}=K_{x_n}B+K\big[\sigma A^{n},B\big].\label{K-tx}
\end{gather}
Accordingly,
\begin{gather*}
\overline\partial K_{t_mt_n}=K_{t_mt_n}B+K_{t_n}\big[A^{m},B\big]+K_{t_m}\big[A^{n},B\big]+K\big[A_m,\big[A^{n},B\big]\big]
\end{gather*}
thus, taking into account the commutativity of $A^m$ and $A^n$, we get $\overline\partial(K_{t_mt_n}-K_{t_nt_m})=(K_{t_mt_n}-K_{t_nt_m})B$ by (\ref{dbar1}). Thus, the commutativity of derivatives
\begin{gather}
K_{t_mt_n}=K_{t_nt_m}\label{K-tt}
\end{gather}
follows due to the unique solvability of the problem (\ref{dbar1}), (\ref{dbar2}). Similarly, we prove that $K_{x_mt_n}=K_{t_nx_m}$ and $K_{x_mx_n}=K_{x_nx_m}$.

In \cite{akp2020}, the time derivatives of the dressing operator for positive times ($n>0$ in ({\ref{comm}})) were calculated in terms of the asymptotic decomposition of the dressing operator $K$
\begin{gather}
	\widetilde{K}(t,x,z)=1+u(t,x)z^{-1}+v(t,x)z^{-2}+w(t,x)z^{-3}+o\big(z^{-3}\big),\label{Kuvw}
\end{gather}
where $u$, $v$, and $w$ are multiplication operators, i.e., their symbols do not depend on $z$. Say, using (\ref{K-tx}) for $n=1$ we get $\overline\partial K_{t_1}=K_{t_1}B+K[A,B]$. This can be written as $\overline\partial(K_{t_1}+KA)=(K_{t_1}+KA)B$, where (\ref{A0}) and (\ref{dbar1}) were used. Due to the condition of unique solvability of~(\ref{dbar1}),~(\ref{dbar2}) we derive that there exists multiplication operator $X$ such that $K_{t_1}+KA=(A+X)K$. Thanks to (\ref{Kuvw}), it is easy to see that it equals to zero, so we have
\begin{gather}
K_{t_1}=[A,K].\label{t1}
\end{gather}
The situation with $K_{x_1}$ is more involved, here analogous multiplication operator does not vanish and by (\ref{dbar2}) we get
\begin{gather}
K_{x_1}=[\sigma A,K]-[\sigma,u]K,\label{Kx1}
\end{gather}
where the multiplication operator $u$ is defined in (\ref{Kuvw}). Combining (\ref{t1}) and (\ref{Kx1}) we get
\begin{gather}
K_{x_1}=\sigma K_{t_1}+[\sigma,K]A-[\sigma,u]K.\label{Kx1t1}
\end{gather}

Our goal here is to extend the approach of \cite{akp2020} to the negative numbers of times in (\ref{comm}). More exactly, we start with the times $t_1$ and $x_1$ as above and we choose either $t_{-1}$ or $x_{-1}$ as the third time according to (\ref{comm22}).

To determine the evolutions with respect to $t_{-1}$ or $x_{-1}$ for the dressing operator we differentiate (\ref{dbar1}) and use (\ref{comm22}):
\begin{gather}
\overline\partial K_{t_{-1}}=K_{t_{-1}}B+K\big[A^{-1},B\big],\qquad \overline\partial K_{x_{-1}}=K_{x_{-1}}B+K\big[\sigma A^{-1},B\big],\label{Ktx1}
\end{gather}
so for the first equality, we have $\overline\partial K_{t_{-1}}=K_{t_{-1}}B+KA^{-1}B-KBA^{-1}$, i.e., thanks to (\ref{A0})
\begin{gather}
\overline\partial (K_{t_{-1}}A+K)=\big(K_{t_{-1}}A+K\big)A^{-1}BA.\label{K}
\end{gather}
We see that situation here is more complicated than in the case of positive numbers of times. There we were able to reduce the equations to the form $\overline\partial(K_{t_n}+KA^n)=(K_{t_n}+KA^{n})B$ due to~(\ref{A0}). While for negative $n$, this equality gives an additional delta-term. Therefore, to use the relation (\ref{K}), we must find replacement for $A^{-1}BA$.

This can be done by introducing a discrete variable, cf.\ \cite{akp2008b} and (\ref{Bt2}) here. We assume that the symbols of $B$, $K$, etc.\
depend on an intermediate variable $n\in\mathbb{Z}$. Denote $\widetilde{B}^{(1)}(t,x,n,z)=\widetilde{B}(t,x,n+1,z)$, $\widetilde{K}^{(1)}(t,x,n,z)=\widetilde{K}(t,x,n+1,z)$ and set
\begin{gather}
B^{(1)}=ABA^{-1}, \qquad B^{(-1)}=A^{-1}BA,\qquad \dots.\label{B1}
\end{gather}
It is easy to see that these shifts commute with times $t$ and $x$: $\big(B^{(1)}\big)_{t_j}=(B_{t_j})^{(1)}$, $\big(B^{(1)}\big)_{x_j}=(B_{x_j})^{(1)}$ and we extend definition of composition law (\ref{comp}) to symbols that depend on $n$ pointwise with respect to this variable. Now $\overline\partial K^{(1)}=K^{(1)}ABA^{-1}$ because of (\ref{dbar1}), so that due to the unique solvability of the problem (\ref{dbar1}), (\ref{dbar2}) there exists a multiplication operator $\psi$ such that
\begin{gather}
K^{(1)}A=(A+\psi)K,\label{K1}
\end{gather}
and thanks to (\ref{Kuvw}) we get
\begin{gather}
\psi=u^{(1)}-u,\label{psi}
\end{gather}
where $u^{(1)}(t,x,n)=u(t,x,n+1)$.
Let us shift $n\to n+1$ of (\ref{K}) that due to (\ref{B1}) gives $\overline\partial \big(K^{(1)}_{t_{-1}}A+K^{(1)}\big)=\big(K^{(1)}_{t_{-1}}A
+K^{(1)}\big)B$ so that, because of (\ref{dbar2}), there exists multiplication operator~$Z$ such that $K^{(1)}_{t_{-1}}A+K^{(1)}=ZK$. Thanks to (\ref{Kuvw}), we get that $Z=1+u^{(1)}_{t_{-1}}$.

It looks like we have constructed a $(3+1)$-dimensional integrable system with the independent variables $t_1$, $x_1$, $t_{-1}$, and $n$. But in fact, we have two different systems here: $t_1$, $x_1$, $n$ (see (\ref{K1})) and $t_1$, $x_1$, $t_{-1}$, because the dependence on $n$ can be excluded. Indeed, substituting $K^{(1)}$ for $K$ by means of (\ref{K1}) and using $\psi$ as new dependent variable in (\ref{psi}) instead of $u^{(1)}$, we get
\begin{gather}
K_{t_1t_{-1}}+K_{t_1}A^{-1}+K_{t_{-1}}A+\psi\big(K_{t_{-1}}+KA^{-1}\big)-u_{t_1}K=0,\label{Ktt}
\\
K_{t_1x_{-1}}+K_{t_1}\sigma A^{-1}+K_{x_{-1}}A+K\sigma+\psi\big(K_{x_{-1}}+K\sigma A^{-1}\big)
-(\sigma+u_{x_{-1}})K=0.
\label{Ktx}
\end{gather}
Here the equation (\ref{Ktx}) is derived by analogy using the second equality in (\ref{Ktx1}). The compa\-ti\-bi\-lity of any of these equations with (\ref{Kx1t1}) can be proved like in (\ref{K-tt}).

Compatible evolutions (\ref{Kx1t1}) and (\ref{Ktt}) or (\ref{Kx1t1}) and (\ref{Ktx}) admit higher (in fact, lower) versions that involve the times $t_{-n}$ and $x_{-n}$, $n>1$, see (\ref{comm3}). By analogy with (\ref{Ktx1}), we get for this case by (\ref{comm3})
\begin{gather}
\overline\partial K_{t_{-n}}=K_{t_{-n}}B+K\big[A^{-n},B\big].\label{n1}
\end{gather}
Multiplying this equality by $A^{n}$ from the right, we use $n$-multiple application of (\ref{B1}): $B^{[-n]}=A^{-n}BA^{n}$. Thus (\ref{n1}) takes the form
\begin{gather*}
\overline\partial\big(K_{t_{-n}}A^{n}+K\big)=\big(K_{t_{-n}}A^{n}+K\big)B^{[-n]},
\end{gather*}
cf.\ (\ref{K}). Again thanks to the assumed unique solvability of the Inverse problem (\ref{dbar1}), (\ref{dbar2}) we get that there exist multiplication
operators $\alpha_0,\dots,\alpha_{n-1}$ such that
\begin{gather}
K^{[n]}_{t_{-n}}A^{n}+K^{[n]}=\sum_{j=0}^{n-1}\alpha_{j}A^{j}K,\label{n3}
\end{gather}
where we applied $n$-fold shift operation. The operators $\alpha_{j}$ are defined in terms of operators $u$, $v$, $w$, etc.\ in (\ref{Kuvw}). We omit these calculations here.

Next, we execute an $(n-1)$-fold shift of a discrete variable in equation (\ref{K1}), which gives
\begin{gather}
K^{[n]}A^{n}=\big(A+\psi^{[n-1]}\big)\big(A+\psi^{[n-2]}\big)\cdots\big(A+\psi\big)K,\label{n4}
\end{gather}
where the multiplication operator $\psi$ was defined in (\ref{psi}). The final expression follows as a~result of inserting of $K^{[n]}$ from (\ref{n4}) to (\ref{n3}), which again cancels dependence on the auxiliary variable $n$. The consideration of dependence on $x_{-n}$ is similar.

\section{Lax pair and nonlinear equations}\label{sec5}

In (\ref{K-tt}), we proved that the commutativity of evolutions (\ref{t1}), (\ref{Kx1}), (\ref{Ktx1}), and (\ref{K1}) is a~direct consequence of commutativity of
evolutions (\ref{comm2}) and (\ref{B1}) and the consequence of the unique solvability of the problem (\ref{dbar1}), (\ref{dbar2}). These conditions lead to the compatibility of the equation (\ref{Kx1t1}) with (\ref{Ktt}), or (\ref{Ktx}), which give nonlinear equations of motion. To simplify these equations, it is reasonable to rewrite them in terms of the Jost solutions defined by means of the symbol of dressing operator:
\begin{gather}
\varphi(t,x,z)=\widetilde{K}(t,x,z){\rm e}^{zt_1+\sigma zx_1+z^{-1}t_{-1}+\sigma z^{-1}x_{-1}}.\label{vhi}
\end{gather}
Here we omit dependence on the discrete variable $n$, since it was excluded from (\ref{Ktt}) and (\ref{Ktx}).

Thanks to this substitution coefficients of the equations (\ref{Kx1t1}), (\ref{Ktt}), and (\ref{Ktx}) become independent on $z$:
\begin{gather}
\varphi_{x_1}-\sigma\varphi_{t_1}+[\sigma,u]\varphi=0,
\label{vhi1}	
\\
\varphi_{t_1t_{-1}}+\psi\varphi_{t_{-1}}-(1+u_{t_{-1}})\varphi=0,
\label{vhitt}
\\
\varphi_{t_1x_{-1}}+\psi\varphi_{x_{-1}}-(\sigma+u_{x_{-1}})\varphi=0,
\label{vhitx}
\end{gather}
where the first equation is the famous two-dimensional linear Zakharov--Shabat problem.

One can also rewrite (\ref{dbar1}) in terms of the Jost solutions. Say, by means of (\ref{Bf1}) we get
\begin{gather}
\dfrac{\partial\varphi(t,x,z)}{\partial\overline z}=\varphi(t,x,\overline{z})g(z),\label{dvhi1}
\end{gather}
and by means of (\ref{Bf2})
\begin{gather}
\dfrac{\partial\varphi(t,x,z)}{\partial\overline z}=\delta(z_{\Re})\int {\rm d}p\varphi(t,x,{\rm i}p)h(p-z_{\Im},z_{\Im}).\label{dvhi2}
\end{gather}
We see that the equations on the Jost solutions are independent on all ``time'' variables $t$ and~$x$. The dependence on them, as well as on
$z$ in (\ref{vhi1})--(\ref{vhitx}) is given by (\ref{dbar2}), which, thanks to~(\ref{vhi}), takes the form
\begin{gather}
\lim_{z\to\infty}\varphi(t,x,z){\rm e}^{-zt_1-\sigma zx_1-z^{-1}t_{-1}-\sigma z^{-1}x_{-1}}=1.\label{vhi3}
\end{gather}
Note that (\ref{dvhi1}) is a standard $\overline\partial$-problem with the normalization condition (\ref{vhi3}), where we must perform
substitution mentioned in (\ref{bf1}). At the same time, (\ref{dvhi2}) shows that the Jost solution in this case is analytic in the left and right half planes of $z$ with discontinuity on the imaginary axis. Thus here inverse problem is given in terms of the Riemann--Hilbert problem, i.e., we define the boundary values of the Jost solution as $\varphi^{\pm}(t,x,{\rm i}z_{\Im})=\lim\limits_{z_{\Re}\to\pm0}\varphi(t,x,z)$ and set
\begin{gather*}
\varphi^{+}(t,x,{\rm i}z_{\Im})-\varphi^{-}(t,x,{\rm i}z_{\Im})=\int {\rm d}p\varphi^{-}(t,x,{\rm i}p)h(p-z_{\Im},z_{\Im}),
\end{gather*}
under the condition (\ref{vhi3}) and substitution given in (\ref{bf2}). The difference of these two formulations of the inverse problem results from the condition of boundedness of the symbol of operator $B$ in (\ref{Bf1}) and (\ref{Bf2}). In the case of (\ref{dvhi1}) $t_n$ are real and $x_n$ are pure imaginary, while in the case of (\ref{dvhi2}) $t_{n}$ and $x_n$ with odd $n$ are real and are pure imaginary for even $n$.

The compatibility of (\ref{vhi1}) with (\ref{vhitt}) and (\ref{vhitx}) follows from (\ref{K-tt}) and (\ref{vhi}). Thus, we get the following theorem.
\begin{Theorem}\qquad

\begin{itemize}\itemsep=0pt
\item[$(a)$] Let the minimal subset of independent variables includes $t_1$, $x_1$, and $t_{-1}$. Then compatibility of \eqref{vhi1} and \eqref{vhitt} gives
\begin{gather}
u_{t_1t_{-1}}\sigma-u_{x_1t_{-1}}-[\sigma,\psi(1+u_{t_{-1}})]+[u_{t_{-1}},[\sigma,u]]=0,
\label{eq1}
\\
\psi_{x_1}-\sigma\psi_{t_1}-[\sigma,u_{t_1}]+[\sigma,\psi]\psi+[[\sigma,u],\psi]=0.
\label{eq2}
\end{gather}
\item[$(b)$] Let the minimal subset of independent variables includes $t_1$, $x_1$, and $x_{-1}$. Then compatibility of \eqref{vhi1} and \eqref{vhitx} gives
\begin{gather}
u_{t_1x_{-1}}\sigma-u_{x_1x_{-1}}-[\sigma,\psi(\sigma+u_{x_{-1}})]+[\sigma+u_{x_{-1}},[\sigma,u]]=0,\label{eq3}
\end{gather}
where the equation for $\psi$ coincides with \eqref{eq2}.
\end{itemize}
\end{Theorem}

It is natural to decompose both matrices $u$ and $\psi$ into diagonal and anti-diagonal parts:
\begin{gather*}
u=u^{\text{d}}+u^{\text{a}},\qquad\psi=\psi^{\text{d}}+\psi^{\text{a}},
\end{gather*}
so $\big[\sigma,u^{\text{d}}\big]=0$, $[\sigma,u^{\text{a}}]=2\sigma u^{\text{a}}$ thanks to (\ref{init}). Then anti-diagonal parts of the equations (\ref{eq1}) and (\ref{eq3}) give
\begin{gather}
u^{\text{a}}_{t_{-1}t_1}+\sigma u^{\text{a}}_{t_{-1}x_1}+2\psi^{\text{a}}\big(1+u^{\text{d}}_{t_{-1}}\big)
+2\psi^{\text{d}}u^{\text{a}}_{t_{-1}}+2\big[u^{\text{a}},u^{\text{d}}_{t_{-1}}\big]=0,
\label{eq1'}
\\[.5ex]
u^{\text{a}}_{x_{-1}t_1}+\sigma u^{\text{a}}_{x_{-1}x_1}+2\psi^{\text{a}}\big(\sigma+u^{\text{d}}_{x_{-1}}\big)
+2\psi^{\text{d}}u^{\text{a}}_{x_{-1}}
-4\sigma u^{\text{a}}+2\big[u^{\text{a}},u^{\text{d}}_{x_{-1}}\big]=0,
\label{eq3'}
\end{gather}
while their diagonal parts reduces to the derivative of (\ref{eq1}) with respect to $t_{-1}$ and of (\ref{eq3}) with respect to $x_{-1}$ of one and the same equation
\begin{gather}
u^{\text{d}}_{t_{1}}-\sigma u^{\text{d}}_{x_{1}}-2\big(u^{\text{a}}\big)^{2}=0\label{eq13}
\end{gather}
that we have integrated here with respect to $t_{-1}$ (or, correspondingly, to $x_{-1}$) under the assumption of the rapid decay of $u$ as $(t_1,x_1)\to\infty$.

Similarly, we derive that diagonal and anti-diagonal parts of (\ref{eq2}) obey
\begin{gather}
\psi^{\text{d}}_{t_1} -\sigma\psi^{\text{d}}_{x_1}-2\sigma\big(\psi^{\text{a}}\big)^{2}+2\big\{u^{\text{a}},\psi^{\text{a}}\big\}=0,
\label{eq2d}
\\
\psi^{\text{a}}_{t_1}-\sigma\psi^{\text{a}}_{x_1}+2u^{\text{a}}_{t_1}
-2\psi^{\text{a}}\psi^{\text{d}}+2\big[\psi^{\text{d}},u^{\text{a}}\big]=0,
\label {eq2a}
\end{gather}
where $\{\cdot,\cdot\}$ denotes anticommutator.
{\sloppy\begin{Corollary}
Each of these systems of four equations, \eqref{eq1'}, \eqref{eq13}, \eqref{eq2d}, \eqref{eq2a}, and~\eqref{eq3'}, \eqref{eq13}, \eqref{eq2d}, \eqref{eq2a}, has only one evolution equation specific to one or another system. The other three equations play an auxiliary role and coincide.
\end{Corollary}}

\section{Dimensional reductions}\label{sec6}
Here we introduce $(1+1)$-dimensional reductions of $(2+1)$-dimensional nonlinear integrable equations constructed above. Such reductions follow due to time evolutions (\ref{Bf1}), (\ref{Bf2}), which, due to (\ref{dbar1}) and (\ref{dbar2}), lead to the same reductions of the dressing operator $K$, and then to reductions of all coefficients of the series (\ref{Kuvw}). The reduction of time dependence of the ope\-ra\-tor~$B$, in turn, is the result of conditions on the supports of the functions $g(z)$ and $h(p,z_{\Im})$ in~(\ref{Bf1}) and (\ref{Bf2}), which reduce the number of independent time variables. For example, for the operator $\widetilde{B}(t,x,z)$ in (\ref{Bf1}), depending on times $t_1$, $x_1$, and $t_{-1}$, we can cancel dependence on $x_1$ by imposing condition
\begin{gather*}
g(z)=\delta(z_{\Re})G(z_{\Im}).
\end{gather*}
Thanks to (\ref{Bf1}), this gives
\begin{gather}
\widetilde{B}(t,x,z)=\exp\biggl({-}2{\rm i}\biggl[z_{\Im}t_1-\dfrac{t_{-1}}{z_{\Im}}\biggr]\biggr)
\delta(z_{\Re})G(z_{\Im}).
\label{s1-2}
\end{gather}
It is clear that this dependence on two variables is preserved in evolution and that thanks to the $\overline\partial$-problem (\ref{dbar1}) and (\ref{dbar2}) and the composition law (\ref{comp}) (or due to (\ref{Bf1expl})) we get the symbol of the operator $K$ also independent of $x_1$. Moreover, this operator is now analytic function for $z_{\Re}\neq0$. Taking into account the independence of the operator $K$ from $x_1$, we must change the definition of the Jost solution, cf.\ (\ref{vhi}),
\begin{gather*}
\varphi(t_1,t_{-1},z)=\widetilde{K}(t,x,z){\rm e}^{zt_1+z^{-1}t_{-1}},%\label{vhi11}
\end{gather*}
so thanks to (\ref{vhi1}) and (\ref{vhitt}), the Jost solution obeys a Lax pair, where the first equation reads~as
\begin{gather}
\sigma\varphi_{t_1}-[\sigma,u]\varphi=z\varphi\sigma,\label{vhi10}
\end{gather}
(cf.\ (\ref{vhi1})) and the second equation coincides with (\ref{vhitt}).

In the same way, we derive compatibility conditions for these equations from (\ref{eq1}) and (\ref{eq2}):
\begin{gather*}
u_{t_1t_{-1}}\sigma-[\sigma,\psi(1+u_{t_{-1}})]+[u_{t_{-1}},[\sigma,u]]=0,%\label{eq10}
\\[.5ex]
\sigma\psi_{t_1}+[\sigma,u_{t_1}]-[\sigma,\psi]\psi-[[\sigma,u],\psi]=0.%\label{eq20}
\end{gather*}
We see that the $\overline\partial$-problem in this case is the Riemann--Hilbert problem for a function analytic in the right and left half planes on the complex $z$-plane with discontinuity given by (\ref{s1-2}) on the imaginary axis. The function $K$ is normalized by the condition (\ref{dbar2}) at $z\to\infty$. Summari\-zing,~(\ref{vhi10}) is nothing but Zakharov--Shabat linear problem \cite{ZSh} that has been extensively studied in the literature, e.g.,~\cite{FT}.

This is not the only reduction applicable to (\ref{Bf1}). Setting there
\begin{gather*}
g(z)=\delta(|z|-1)g(z_{\Im}),
\end{gather*}
we get the scattering data, i.e., symbol of operator $B$, depending on two variables $t_1-t_{-1}$ and~$x_1$:
\begin{gather}
\widetilde{B}(t,x,z)=\delta(|z|-1)\exp\bigl(-2{\rm i}z_{\Im}(t_1-t_{-1})+2\sigma z_{\Re}x_1\bigr)g(z_{\Im}).\label{Bf3}
 \end{gather}
Thus after shifting $t_1\to t_1+t_{-1}$, we exclude the dependence on $t_{-1}$ from $B$, and then from~$K$. Now, because of the delta-function in (\ref{Bf3}), we reduce the inverse problem (\ref{dbar1}) to the Riemann--Hilbert problem on the circle $|z|=1$ and the normalization condition (\ref{dbar2}). Now
we define the Jost solution by means of the relation
\begin{gather*}
\varphi(t_1,x_1,z)=\widetilde{K}(t_1+t_{-1},t_{-1},x_1,z){\rm e}^{zt_1+\sigma zx_{1}},
\end{gather*}
where the r.h.s.\ does not depend on $t_{-1}$. The integrable equation follows from (\ref{eq1}):
\begin{gather*}
u_{t_1t_{1}}\sigma-u_{x_1t_{1}}-[\sigma,\psi(1+u_{t_{1}})]+[u_{t_{1}},[\sigma,u]]=0,%\label{eq4}
\end{gather*}
where the second equation (\ref{eq2}) is left unchanged.

By analogy, we can consider the reductions of the symbol of operator $B$ in (\ref{Bf2expl}), i.e., when~$t_1$,~$x_1$, and $x_{-1}$ are chosen as independent variables.

\section{Concluding remarks}\label{sec7}
In the above derivation of nonlinear integrable equations we needed some essential assumptions, the main was the condition of unique solvability of the $\overline\partial$-problem (\ref{dbar1}), (\ref{dbar2}). But when nonlinear equation is derived, these assumptions are not necessary: the nonlinear equation is given as a compatibility condition of a Lax pair. On the other hand, the existence of linear equations given by commutator identities always leads to nonlinear integrable equations, as was shown above.

 \subsection*{Acknowledgements}
 This research is supported by a grant from the Russian Science Foundation (Project No.~19-11-00062).

\pdfbookmark[1]{References}{ref}
\LastPageEnding

\end{document}